\documentclass[traditabstract]{aa}
\usepackage{graphicx}
\usepackage{txfonts}
\usepackage{natbib}
\bibpunct{(}{)}{;}{a}{}{,}
\usepackage{journals}
\usepackage{multirow}
\usepackage{url}

\usepackage{color}

\begin{document}

\title{Virtual Observatory based
identification of AX~J194939+2631 as a new cataclysmic variable\thanks{Based
on observations collected at the German-Spanish Astronomical Center, Calar
Alto, jointly operated by the Max-Planck-Institut f\"ur Astronomie
Heidelberg and the Instituto de Astrof\'{\i}sica de Andaluc\'{\i}a (CSIC).}}
\titlerunning{A new CV among unidentified ASCA sources}

\author{Ivan Yu. Zolotukhin\inst{1,2} \and Igor V. Chilingarian\inst{3,2}}
\authorrunning{Zolotukhin \& Chilingarian}

\offprints{Ivan Zolotukhin, \email{iz@sai.msu.ru}}

\institute{
Observatoire de Paris-Meudon, LERMA, UMR~8112, 61 Av. de l'Observatoire, 75014 Paris, France
\and
Sternberg Astronomical Institute, Moscow State University, Universitetskij pr., 13, 119992, Moscow, Russia
\and
Observatoire astronomique de Strasbourg, UMR 7550, Universit\'e de Strasbourg/CNRS, 11 rue de l'Universit\'e, 67000 Strasbourg, France
}

\date{Received August 12, 2010; Accepted October 11, 2010}

\abstract{
We report the discovery of a new cataclysmic variable (CV) among
unidentified objects from the {\it ASCA} Galactic Plane Survey
made using the Virtual Observatory data mining.
First, we identified \object{AX~J194939+2631} with IPHAS
J194938.39+263149.2, the only prominent H$\alpha$ emitter among 400
sources in a 1~arcmin field of the IPHAS survey, then secured as a single 
faint X-ray source found in an archival {\it Chandra} dataset. Spectroscopic follow-up
with the 3.5-m Calar Alto telescope confirmed its classification as a CV,
possibly of magnetic nature. Our analysis suggests that
AX~J194939+2631 is a medium distance system ($d \approx 0.6$~kpc) containing
a late-$K$ or early-$M$ type dwarf as a secondary component and a partially
disrupted accretion disc revealed by the double-peaked H$\alpha$ line. 
However, additional deep observations are needed to confirm our tentative 
classification of this object as an intermediate polar.
}


\maketitle

\section{Introduction}

Cataclysmic variables are interacting binary systems containing a white
dwarf (WD) primary star and a low-mass secondary, usually late-type main
sequence star. Moving about the primary star with a typical orbital period of
1 day, the secondary star fills its Roche lobe and experiences the mass
loss. This matter then spirals down to the non-magnetic WD, forming an
accretion disc. These systems are roughly divided into 3 main subtypes:
novae (CVs exhibiting thermonuclear bursts on a WD surface resulting in a
6--19~mag increase of their optical luminosities), dwarf novae (CVs where pure
disc instability causes quasiperiodic outbursts that temporarily increase
optical fluxes by 2--8 mag), and nova-like CVs (non-eruptive CVs
characterized by an approximately constant, high rate mass transfer, a
prominent accretion disc, and a high luminosity). Magnetic CVs by definition
are those, where magnetic field of the primary companion disrupts the
accretion disc, either partially (intermediate polars) or totally (polars).
The comprehensive and ultimate overview of CV subclasses and their
properties are given in \citet{warner95}.

Over recent years, numerous authors undertook comprehensive CV
searches in large publicly available surveys
\citep[e.g.][]{szkody02,witham07,witham08,denisenko10}. New generation H$\alpha$
surveys are especially convenient for studies like these. The recent Isaac Newton 
Telescope (INT) Photometric H$\alpha$
Survey of the northern Galactic plane (IPHAS) covers $-5^\circ < b <
+5^\circ$ latitude range providing two broad-band SDSS $r'$ and $i'$ and a
narrow-band H$\alpha$ magnitudes for about 300 million sources up to $r'
\sim 20$~mag. A detailed introduction to the survey is given in
\citep{drew05}.

Using the two-colour diagram $(r' - i')$ vs. $(r' - H\alpha)$ as a primary
analysis method, \citet{witham06} reported 70~per~cent recovery rate of
known CVs in IPHAS, because most of them demonstrate H$\alpha$ emission steadily
detectable in the IPHAS photometric system. Hence, this method
provides a highly efficient selection criterion to distinguish between CV
candidates and field stars.

\section{Identification of AX~J194939+2631}

Generally interested in optical/NIR identification of X-ray sources 
with arcsec-scale error boxes \citep{zolotukhin10a,zolotukhin10b}, 
we decided to employ the \citet{witham06} approach in attempt to 
identify CVs among unidentified X-ray sources with challenging 
positional uncertainties, unlikely to be studied by conventional 
identification techniques.

We use the {\it ASCA} Galactic Plane Survey \citep{sugizaki01}
carried out in 1996--1999 in a stripe $|l| \lesssim 45^\circ, |b| \lesssim
0.4^\circ$ partially overlapping the IPHAS footprint. This survey,
originally intended to study the extended Galactic ridge X-ray emission
recently explained by \citet{revnivtsev09}, produced 163 point sources
brighter than $10^{-12.5}$~erg~cm$^{-2}$~s$^{-1}$ in the 0.7--10~keV range, 107
of which remained unidentified. Since then, some authors attempted to
explore unidentified {\it ASCA} sources using dedicated observations (see e.g.
\citet{kaur10}), although those studies were hampered by large
source density and high probability of random coincidence in 1~arcmin-scale
X-ray error boxes of {\it ASCA} detections in the Galactic plane. 

In our study we decided to adopt the innovative VO-powered research
concept \citep{chilingarian09a} including the three steps: (1) search for
candidate object(s) using VO technologies and infrastructure in a
semi-automatic way; (2) follow-up observations; (3) interpretation of newly
obtained data together with all the information available from the VO. Worth
mentioning, that the VO-based identification of an X-ray source presented
hereafter was made during the tutorial for undergraduate students held at
the Sternberg Astronomical Institute in May 2009\footnote{The complete
step-by-step tutorial description is published at the Euro-VO web-site at
\url{http://www.euro-vo.org/pub/fc/workflows.html}}.

As we were to search CVs among unidentified {\it ASCA} sources,
we first maximized the corresponding probability by preselection of those (1)
falling into the IPHAS Initial Data Release (IDR) footprint (33 sources) (2)
having hard X-ray spectra (spectral photon index $\Gamma < 3$) and (3) high
estimated column densities ($N_H > 10^{22}$~cm$^{-2}$) indicative of large
distances from the Sun. Thus we obtained a list of 6 sources which were not
expected to be coronal active stars due to the imposed criteria. Then we
manually explored colour--colour diagrams of IPHAS datasets in 1.3~arcmin
circular fields around {\it ASCA} coordinates of these sources using VO
tools, namely {\sc cds aladin} \citep{bonnarel00} and {\sc topcat}
\citep{taylor05} for quick-look purposes. We somewhat arbitrarily increased
the {\it ASCA} error box radius from 1~arcmin uncertainty declared by
\citet{sugizaki01} to 1.3~arcmin due to our own considerations. On average,
there are $\sim 400$ IPHAS sources falling inside such a positional uncertainty.

\begin{figure}
\resizebox{\hsize}{!}{\includegraphics{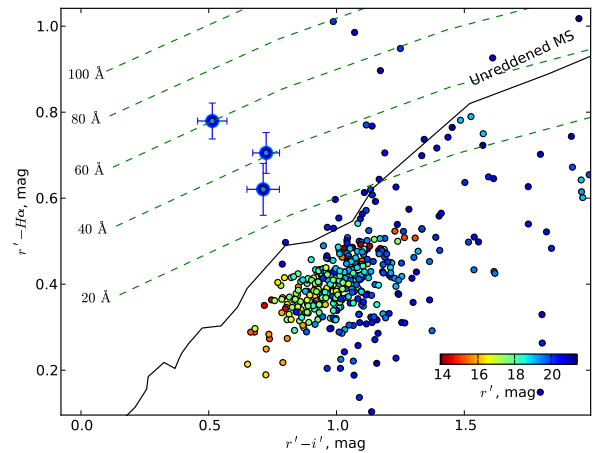}}
\caption{Colour-colour diagram of the IPHAS detections (not sources) in the
 1.3~arcmin field around ASCA coordinates of AX J194939+2631. Objects with
 H$\alpha$ excess are located towards the top of the diagram. The $r'$ magnitude
 is colour-coded. Black solid line indicates unreddened main sequence, while
 green dashed lines are those of constant H$\alpha$ EW with corresponding
 values labelled on the left (from \citet{drew05}). Three detections of
 IPHAS J194938.39+263149.2 marked with larger circles correspond to the only
 prominent H$\alpha$ emitter in the field. Error bars given for it are
 applicable to other detections as well.} \label{color_color}
\end{figure}

\begin{figure}
\includegraphics[width=\hsize]{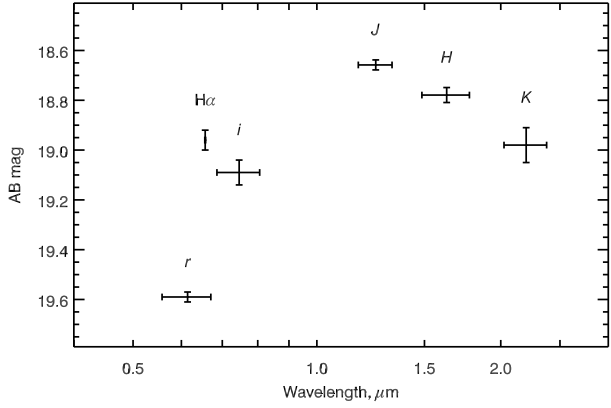}
\caption{The reconstructed optical/NIR SED of the source.}
 \label{iphas_ukidss_SED}
\end{figure}

For one of the sources, AX~J194939+2631, we found single easily detectable
prominent H$\alpha$ emitter in IPHAS data (see Fig.~\ref{color_color}).
Despite of the presence of other sources with signs of the H$\alpha$
emission excess, IPHAS~J194938.39+263149.2 is the only one in the field
exhibiting permanent excess of high enough confidence level (see
typical errors for its magnitude in Fig.~\ref{color_color}). The photometric
data gave the H$\alpha$ EW estimate of 30--60~\AA, quite typical for a CV. As
the candidate falls to the corners of individual IPHAS CCD frames, it was
observed 3 times on Jul 14, 2005 from 01:04 UT to 02:24 UT, giving clear
signs of 0.2~mag variability in $i'$ and H$\alpha$ (see
Table~\ref{main_table}). The source is also present in UKIRT Infrared Deep
Sky Survey (UKIDSS) Data Release 4 data \citep{lawrence07}. This way, using
only VO resources, we constructed its optical/NIR spectral energy
distribution (SED) shown in Fig.~\ref{iphas_ukidss_SED}. Additional important 
information on the source was also found in the USNO-B1.0 catalogue \citep{monet03}.

Based on these data, we considered IPHAS~J194938.39+263149.2 to be a
tentative counterpart of AX~J194939+2631 and attempted to get more evidences
of their probable association.

\section{Additional data}

\begin{figure}
\includegraphics[width=\hsize]{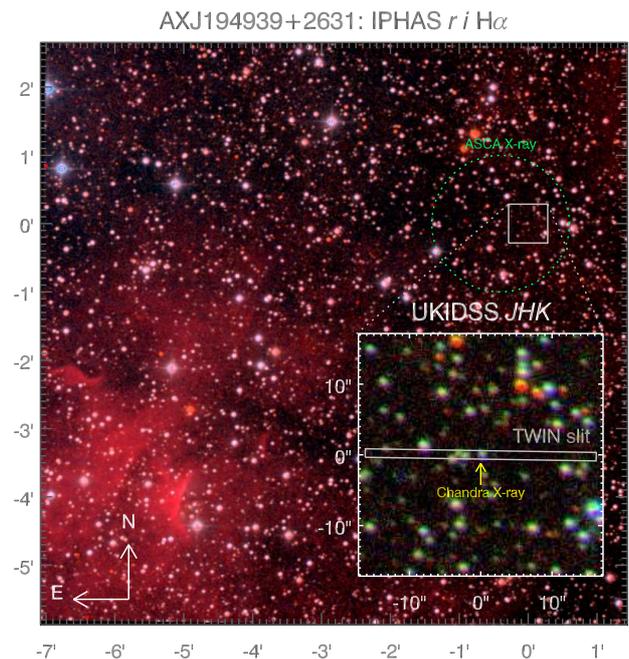}
\caption{Data for AX J194939+2631 available in the Virtual Observatory.
The background image is a false colour composite made of $r\prime$ and
$i\prime$ IPHAS images with H$\alpha$ added to the red channel using the
\citet{lupton04} algorithm. The {\it
ASCA} positional uncertainty is shown as a green circle. A
30$\times$30~arcsec fragment of the error circle containing the
IPHAS J194938.39+263149.2 source as seen in the near-infrared UKIDSS data 
is displayed in the bottom right inset. The {\it Chandra} X-ray
source position is shown by an arrow, as well as the slit of the Calar Alto 
TWIN spectrograph (see Section~3.2).
}
 \label{iphas_ukidss_field}
\end{figure}

\subsection{Chandra X-ray data}

Having performed the preliminary VO-based identification with the most
prominent H$\alpha$ emitter in the field, we found and analysed yet
unpublished publicly available X-ray data of this region obtained with the
{\it Chandra} satellite on Jan 8, 2008 in the framework of the
\citet{murray06} programme (dataset ID 8236).

\begin{figure}
\resizebox{\hsize}{!}{\includegraphics[angle=270]{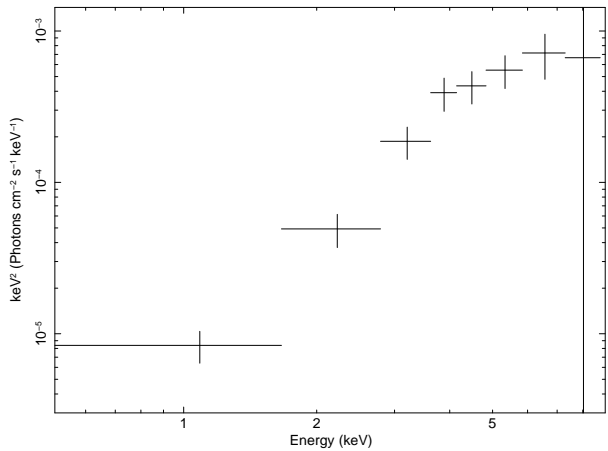}}
\caption{ X-ray spectrum of AX J194939+2631 obtained in a 1.74~ksec exposure 
with {\it Chandra}/ACIS detector in a frame of \citet{murray06} programme. }
 \label{chandra_spectrum}
\end{figure}

The {\it Chandra}/ACIS X-ray image taken with 1.74~ksec exposure contains
a single faint object within the {\it ASCA} error box at the position
RA=19:49:38.39, Dec=26:31:49.1 (J2000) determined with the 0.6~arcsec
uncertainty (90~per~cent confidence), that agrees well with the optical
position of IPHAS J194938.39+263149.2. The X-ray dataset contains only 116
photons. Translated into 0.5-10~keV flux of $1.05 \times
10^{-12}$~erg~cm$^{-2}$~s$^{-1}$, it does not allow us to characterize its
spectrum reliably (see Fig.~\ref{chandra_spectrum}). The best-fitting model
has a spectral photon index of 1.5 with column density 
$N_H \simeq 5.5 \pm 0.5 \times 10^{22}$~cm$^-2$.

Therefore, the {\it Chandra} data allowed us to confirm the association of a
prominent H$\alpha$ emitter with AX J194939+2631, corroborating {\it ASCA}
estimates of $N_H$ and an X-ray flux. Hence, after careful inspection of VO
resources we obtained sufficient information for reliable follow-up studies
by dedicated observations (see Table~\ref{main_table}). At that stage it
already became clear, that the source was either a CV, or a high-mass X-ray
binary (HMXB), based on its photometric H$\alpha$ EW estimate and hard X-ray
spectrum.

\begin{table}
\caption{Summary of the data on AX J194939+2631, immediately available in
the VO after association with IPHAS J194938.39+263149.2 source without
additional observations. IPHAS magnitudes are given for 3 consequent epochs
(see text).\label{main_table}}
\begin{tabular}{rl}
\hline
\hline
\multicolumn{2}{c}{{\it ASCA}} \\
\hline
Name & AX J194939+2631 \\
($l$, $b$) & ($62.937^\circ$, $0.203^\circ$) \\
Flux, 0.7--10 keV & $0.97 \times 10^{-12}$~erg~cm$^{-2}$~s$^{-1}$ \\
Spectral index & $2.61^{+2.43}_{-1.40}$ \\
Column density & $3.39^{+1.99}_{-1.73} \times 10^{22}$~cm$^{-2}$ \\
\hline
\hline
\multicolumn{2}{c}{IPHAS and UKIDSS} \\
\hline
Name & IPHAS J194938.39+263149.2 \\
R.A., Dec. (J2000) & ($297.409912^\circ$, $26.530325^\circ$) \\
$r'$ & $19.42 \pm 0.02, 19.48 \pm 0.03, 19.43 \pm 0.02$ \\
$i'$ & $18.69 \pm 0.05, 18.76 \pm 0.06, 18.91 \pm 0.05$ \\
H$\alpha$ & $18.71 \pm 0.04, 18.86 \pm 0.05, 18.65 \pm 0.03$ \\
H$\alpha$ EW estimate & variable, 30--60 \AA \\
$J$ & $17.72 \pm 0.02$ \\
$H$ & $17.40 \pm 0.03$ \\
$K$ & $17.08 \pm 0.07$ \\
\hline
\hline
\multicolumn{2}{c}{USNO-B1.0} \\
\hline
Name & USNO-B1.0 1165-0456015 \\
$B1$ & 20.04 mag \\
$B2$ & 20.20 mag \\
$R1$ & 20.12 mag \\
$R2$ & 19.42 mag \\
proper motion, RA & $-8 \pm 1$ mas/year \\
proper motion, Dec & $-4 \pm 11$ mas/year \\
\hline
\hline
\multicolumn{2}{c}{{\it Chandra}} \\
\hline
Flux, 0.5--10 keV & $1.05 \times 10^{-12}$~erg~cm$^{-2}$~s$^{-1}$ \\
Column density & $5.5 \pm 0.5 \times 10^{22}$~cm$^{-2}$ \\
Spectral index & 1.5 \\
\hline
\end{tabular}
\end{table}

\subsection{Optical spectroscopic data}

To choose between HMXB and CV we performed dedicated optical spectroscopic follow-up observations.
The data were collected on Jul 12, 2009 in the service mode with
the TWIN long-slit spectrograph mounted at the 3.5~m telescope of the
German--Spanish Astronomical Centre (Calar Alto Observatory) in the
framework of the Director's Discretionary Time proposal ``Optical
spectroscopy of AX~J1949.6+2631'' (P.I.: IZ). The total integration time of
2.5~h was splitted into five 1800~sec-long exposures. The \textit{T08} and
\textit{T10} gratings were used in the blue and red arms of the TWIN
spectrograph simultaneously with the dichroic beam-splitter covering the
wavelength bands 3800--5600~\AA\ and 5550--6900~\AA\ with the spectral
resolving power $R \approx 2500$ and $R \approx 5000$ respectively for a
slit width of 1.2~arcsec. The night time calibration included bias frames,
dome flat fields, He-Ar arc line spectra, and a spectrum of the
\textit{Feige~110} standard star.

We reduced the data using the generic IFU/longslit data reduction pipeline
implemented in {\sc idl} following the same steps as those described in
\citet{chilingarian09c}. The only significant difference was the sky
subtraction procedure. The TWIN data in our configuration are moderately
undersampled in the blue arm and strongly undersampled in the red arm (FWHM
$\sim$1~pix). Therefore, in order to prevent artifacts originating from the
interpolation of undersampled air glow lines, we used the sky subtraction
technique proposed by \citet{kelson03}. We used the entire slit length of
6~arcmin to estimate the oversampled sky spectrum, which we then
approximated using the 4-th order $b$-splines with equidistant nodes every
0.3~\AA. This procedure resulted in a nearly Poisson sky subtraction
quality.

The source is projected on-to a peripheral part of a H{\sc ii} region well
visible in Fig.~\ref{iphas_ukidss_field}. Therefore, its spectra are 
contaminated with narrow nebular emission lines. Since the nebula surface
brightness changes along the slit, our sky subtraction technique left
significant [positive] residuals at the source position in the
following spectral lines: H$\alpha$, [N{\sc ii}] ($\lambda=6548, 6584$~\AA),
[S{\sc ii}] ($\lambda = 6717, 6731$~\AA). We used two 6~arcsec long segments
taken on both sides of the source, interpolated the line residuals in five
narrow wavelength regions ($\pm3$~\AA) around these lines, and subtracted
them from the source spectrum. The absence of any noticeable residuals in the
final extracted source spectrum at the position of the four narrow forbidden
lines suggests the high reliability of this procedure in the H$\alpha$ line as
well.

Then, the sky subtracted spectra were linearised using the wavelength
solution defined from the automatic identification of the arc line spectra.
The systematic errors of the wavelength solution are about 0.07~\AA\ and
0.05~\AA\ in the blue and red arms respectively.

Interestingly, the residuals of the sky subtraction with the
\citet{kelson03} technique did not exceed the Poisson noise in the blue arm
spectra even in the H$\beta$ and [O{\sc iii}] lines, suggesting high
interstellar extinction and, consequently, a large distance to the H{\sc ii}
region.

\begin{figure*}
\includegraphics[width=0.57\textwidth]{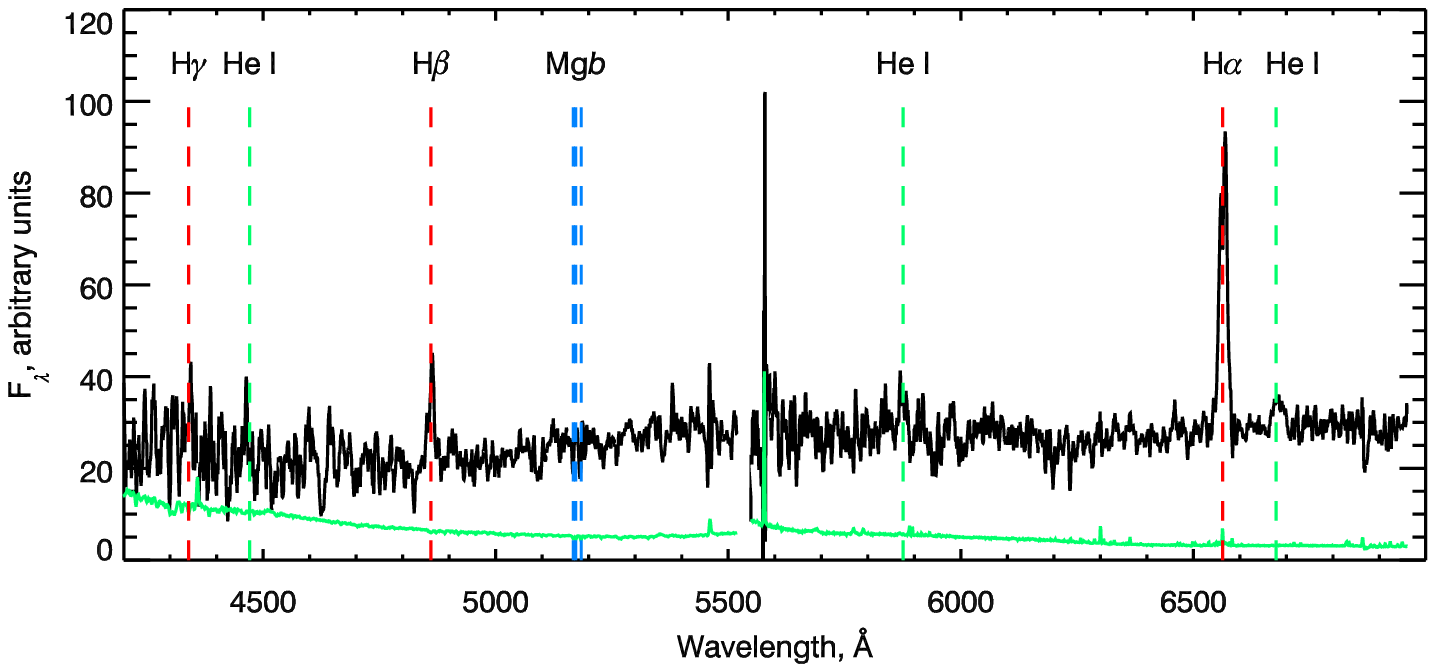}
\includegraphics[width=0.4\textwidth]{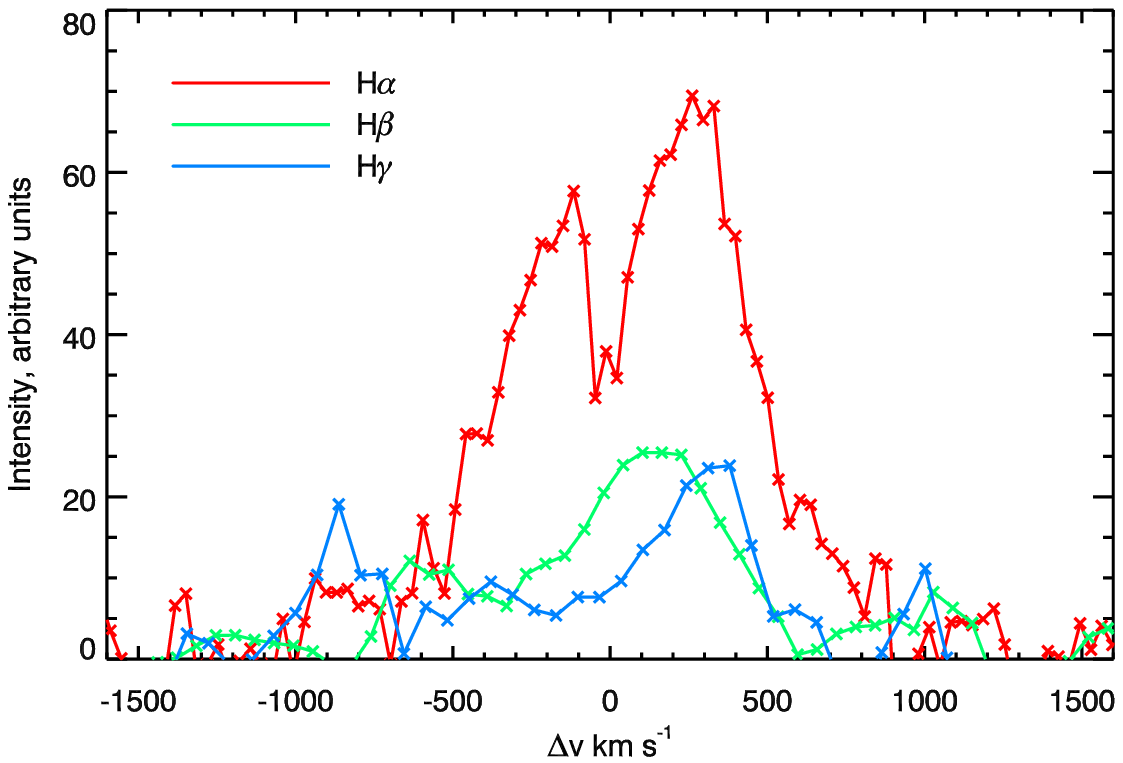}
\caption{{\it Left panel}: the merged optical spectrum of the AX
J194939+2631 counterpart smoothed with a 4~\AA\ box-car for presentation
purposes. {\it Right panel}: continuum subtracted profiles of Balmer
emission lines. The H$\beta$ and H$\gamma$ profiles are smoothed with a
3~\AA\ box-car.}
 \label{spectrum}
\end{figure*}

The co-added, extracted, merged spectrum of IPHAS~J194938.39+263149.2, flux 
calibrated up-to an arbitrary constant is displayed in Fig.~\ref{spectrum}.
Several broad emission lines of neutral hydrogen and helium are clearly visible. 
He{\sc i} line ($\lambda$=4471~\AA) is detected only marginally, as our 
measurements are hampered by the low sensitivity in the blue spectral range.
Where possible, we computed lines fluxes and equivalent widths by estimating the
pseudo-continuum level at the regions beyond 1000~km~s$^{-1}$ from the expected
line central wavelength assuming zero line-of-sight velocities. The obtained
equivalent widths and flux ratios are provided in Table~\ref{ew_table}.

\begin{table}
\centering
\caption{Equivalent widths of detected emission lines and flux ratios
relative to H$\alpha$.}
\label{ew_table}
\begin{tabular}{lrc}
\hline
\hline
Line & EW, \AA & line ratio \\
\hline
H$\alpha$ & $-44.7 \pm 2.5$ & 1.0 \\
H$\beta$  & $-17.3 \pm 3.9$ & 3.5$\pm$1.0 \\
H$\gamma$ & $-10.6 \pm 3.3$ & 5.5$\pm$2.0 \\
He {\sc i} 4471 & $-4.0 \pm 3.2$ & 14$\pm$11 \\
He {\sc i} 5876 & $-5.2 \pm 2.5$ & 8.0$\pm$4.2 \\
He {\sc i} 6678 & $-5.8 \pm 1.9$ & 7.3$\pm$2.8 \\
\hline
\end{tabular}
\end{table}

\section{Discussion}

Detected at their restframe wavelengths, He emission lines and central
H$\alpha$ absorption, as well as the non-zero proper motion of the object in
USNO-B1.0 (at least along one coordinate, see Table~\ref{main_table}),
clearly indicate that this optical source belongs to the Galaxy. The central
absorption in H$\alpha$ has zero velocity within $\pm 50$~km~s$^{-1}$. If the
source had been distant, at this Galactic longitude ($l=62.9$~deg) its
radial velocity would have significant non-zero value due to the Galactic
differential rotation. Sign of its RA proper motion is also in favour of it
to reside in the foreground part of the Galaxy, i.e. much closer than 8~kpc.
Relatively nearby location automatically rules out the high-mass X-ray
binary nature of our object, as its derived X-ray luminosity is too low for
this class of sources: $L_X \simeq 4 \times 10^{33} (d /
8$~kpc$)^2$~erg~cm$^{-2}$.

The observed small Balmer decrement likely means insignificant
interstellar reddening. The assumptions of typical conditions in the emitting
region ($H\alpha$/$H\beta$ flux ratio of 2.86) and the standard extinction
law are translated into $E(B-V) \simeq 0.2$ and $A_V \simeq 0.6$. In this
case, the mismatch with X-ray data showing significant line-of-sight
absorption can be explained by the difference in the geometry of emitting
regions: harder radiation comes from the central source with local
absorption material, whereas the optical spectrum originates from outer
parts of the accretion disc subject to much lower intrinsic extinction. The
circumstellar nature of the X-ray absorption is supported by the fact that
the whole line-of-sight Galactic extinction of $A_V \simeq 21$~mag
\citep{schlegel98} is not enough to account for {\it ASCA} and {\it Chandra}
values of $N_H \sim 3 \dots 5 \times 10^{22}$~cm$^{-2}$, assuming the
standard extinction law with $R=3.1$. The 3D Galaxy extinction map
\citep{marshall06} provides an estimate of $A_V \simeq 2$~mag at 1.5~kpc
distance in this direction, suggesting that the source resides at $d \simeq
500$~pc, if we scale the extinction linearly.

The optical/NIR SED of the source (see Fig.~\ref{iphas_ukidss_SED}) closely matches 
that of an early-M dwarf, with a pronounced blue excess in the $r' - i'$ colour \citep{hewett06},
which actually falls into the most densely populated region in the
colour distribution of known CVs \citep{witham06}. Weak barely detected Mg$b$
absorption together with the lack of strong molecular bands and red
continuum suggests the same classification. Early-M dwarfs are faint objects
with a relatively narrow range of absolute magnitudes ($M_K = 8\dots9$).
Hence, we estimate the distance to the object in a relatively
extinction-free NIR band as $d \simeq 400\dots650$~pc, which is in a good
agreement with the value derived above. The source therefore possesses quite
moderate X-ray luminosity in the 0.7--10~keV range, $L_X \simeq 2 \times
10^{31}$~erg~cm$^{-2}$ at 600~pc, though it is not certain how
the intrinsic extinction can affect the observed value.

We detect hydrogen emission lines up-to H$\gamma$ in the optical
spectrum of the source as well as some He{\sc i} lines, both almost
unambiguously characteristic to CVs. However, similar features can be present
in optical spectra of other object classes such as active late-type stars,
symbiotics and Be stars. Symbiotics do exhibit absorptions of a late-type
giant and significantly broader EW of H$\alpha$ (hundreds of \AA), which we
do not observe here. Be stars can be discerned on a basis of their continua:
they show a pure SED of an early-type star, clearly not the case for our
data. Active late-type stars typically have much narrower H$\alpha$, known
to be less than 10 \AA\ \citep{pettersen89}, as well as strong molecular
absorptions, which we do not detect. We stress that these object types
are not known to produce such hard X-ray spectra as demonstrated in
Fig.~\ref{chandra_spectrum}. Finally, the He{\sc i} ($\lambda=$6678~\AA)
emission line is present in the spectrum, which is not expected for other
types of systems \citep{witham07}. We therefore clearly classify the source
as a CV.

Assuming $(V - r) = 0$, we can estimate the intrinsic $F_X/F_{opt}$ ratio to
be around 10, corrected for intervening Galactic extinction in the optical.
This quite high X-ray to optical ratio, as well as the hard spectrum and
significant intrinsic extinction in X-ray, rule out the \emph{dwarf nova}
classification, suggesting the source to be a \emph{nova-like} variable,
either magnetic or not \citep{warner95}. The double-peaked H$\alpha$ profile
(see Fig.~\ref{spectrum}, right panel) indicates the presence of an
accretion disc, so the polar nature looks unfavourable. However, an
intermediate polar with an accretion ring at inclinations $i \sim
60\dots70$~deg is able to produce such a profile with deep absorption and
moderate separation between the peaks. We cannot explain single-peaked
redshifted profiles of other hydrogen emission lines without risk of
overinterpreting our low signal-to-noise data, but this is possibly caused
by different geometries of emitting regions (though H$\alpha$ peaks are also
asymmetric, the red one being stronger) and a 2.5~h long integration during
the corresponding phase of a spin wave. At the same time, if the orbital
period is comparable to or lower than 2.5~h, this asymmetry can suggest
the emission to originate predominantly from only one pole \citep{warner95,
hellier99}.

We notice, however, that the main difficulty with the magnetic CV explanation
is the lack of high ionisation lines (He{\sc ii}, C{\sc iii}, N{\sc iii}) in
the observed spectrum. Although, this can be explained by the low
signal-to-noise ratio of our optical data.
Additional data of better quality should be obtained to make solid
conclusions about the nature of this system. X-ray timing studies to
possibly detect periodicities and/or phase-resolved deep spectroscopic
observations are required to confirm our tentative classification of this
system as an intermediate polar.

\begin{acknowledgements}
We thank Calar Alto Observatory for allocation of director's discretionary
time to this programme. We would like to thank M.~Revnivtsev for his
invaluable help with processing of the {\it Chandra} dataset. This research has
made use of the VizieR catalogue access tool, CDS, Strasbourg, France. IZ was 
supported by Russian state contract No. 02.740.11.0575 and a grant of VO-Paris 
Data Centre.
\end{acknowledgements}

\bibliographystyle{spphys}
\bibliography{iz_bib}

\end{document}